\documentclass[conference]{IEEEtran}
\usepackage{amsmath,amssymb}
\usepackage{graphicx}
\usepackage{cite}
\usepackage{booktabs}
\usepackage{hyperref}
\usepackage{xcolor}
\usepackage{subcaption}
\usepackage{balance}
\begin{document}
% \IEEEaftertitletext{\vspace{-1mm}}

% ===== Some candidate options for the title (rewording possible) ===== 

% ======= Current option (if EDAS allows this first-line densification hack) ======= Thoughts?
%\title{How to Receive Focused Beams in Near-Field MIMO? Analytical Combiners and Array Size Optimization}
\title{How\hspace{1.45mm}to\hspace{1.45mm}Receive\hspace{1.45mm}Focused\hspace{1.45mm}Beams\hspace{1.45mm}in\hspace{1.45mm}Near-Field\hspace{1.45mm}MIMO? Analytical Combiners and Array Size Optimization\vspace{-2mm}}

\author{
\IEEEauthorblockN{Phinehas Chinthalagattu, Vitaly Petrov, and Emil Björnson}
\IEEEauthorblockA{
Department of Communication Systems, KTH Royal Institute of Technology, Sweden\\
Email: \{phineas, vitalyp, emilbjo\}@kth.se\vspace{-4mm}
}
}

\maketitle

\begin{abstract}
This paper investigates receive combining for near-field beamfocusing in line-of-sight channels with antenna array misalignment. We propose and evaluate two \emph{analytical} combiners: analog ANF--AC and digital ANF--DC. Beyond focusing the beam toward the transmitter, ANF--AC benefits from the proposed phase-flip corrections (PFCs), while ANF--DC additionally exploits Dirichlet-kernel-based amplitude tapering. ANF--AC closely matches theoretically optimal analog equal-gain combining (EGC), with an SNR gap below~0.006\,dB. Similarly, ANF--DC matches optimal digital maximum-ratio combining (MRC). We then leverage the analytical properties of these combiners for link-level evaluation and optimization, deriving a \emph{closed-form expression for the optimal transmitter--receiver array-size trade-off} that maximizes received SNR with analog combining. We also show that the optimized analog receiver performs \emph{within 0.84\,dB} of its fully digital MRC-equivalent counterpart. These results demonstrate that optimized analog combining (with potentially lower implementation complexity and cost, particularly at higher frequencies) can get close to the fully digital performance. Beyond the theoretical insights presented, the analytical tractability and EGC-like / MRC-like performance of our combiners enable \emph{accurate network-level mathematical studies of beamfocusing}, such as coverage, capacity, and interference analysis and optimization toward efficient 6G+ near-field MIMO communication systems.
\end{abstract}

%%%%%%%%%%%%%%%%%%%%%%%%%%%%%%%%%%%%%%%%%%%%%%%%%%%%%%%%%%%%%%%%%%%%%%%%%%%%%%%%%%%%%%%%%%%%%%%%%%%%%%%%%%%%%%%%%%%%%%%%%%%%%%%%%%%%%%%
%%%%%%%%%%%%%%%%%%%%%%%%%%%%%%%%%%%%        Introduction        %%%%%%%%%%%%%%%%%%%%%%%%%%%%%%%%%%%%%%%%%%%%%%%%%%%%%%%%%%%%%%%%%%%%%%%
%%%%%%%%%%%%%%%%%%%%%%%%%%%%%%%%%%%%%%%%%%%%%%%%%%%%%%%%%%%%%%%%%%%%%%%%%%%%%%%%%%%%%%%%%%%%%%%%%%%%%%%%%%%%%%%%%%%%%%%%%%%%%%%%%%%%%%%
\vspace{-5mm}
\section{Introduction}
\label{sec:introduction}
%\vspace{-1mm}
The evolution toward sixth-generation (6G) and beyond wireless systems targets higher data rates and greater network capacity, driving the development of large-aperture antenna arrays and the exploration of higher-frequency bands, including millimeter-wave and terahertz frequencies~\cite{cui2023nearfield,liu2023nearfield}. Meanwhile, large electrical apertures together with short wavelengths extend the radiative near-field (NF) region, where non-planar wavefronts make the channel strongly dependent on both signal direction and receiver distance~\cite{lu2024tutorial}.

Unlike conventional far-field beam steering, which concentrates signals in an angular direction, NF beamfocusing concentrates energy around a target location defined by both angle and range~\cite{kosasih2024finite,zhang2022beam}. This range-dependent spatial selectivity has driven research into NF connectivity using large-scale multiple-input-multiple-output (XL-MIMO) systems~\cite{cui2023nearfield,lu2024tutorial}, NF integrated sensing and communications (ISAC~\cite{sun2024isac}), and even NF wireless power transfer~\cite{zhang2025wpt}. These applications highlight the potential of NF beamfocusing for spatial selectivity.

Most recent studies on NF beamfocusing primarily address transmitter (Tx) design and optimization. On the receiver (Rx) side, two main approaches are commonly adopted. The first is the idealized point-receiver model, which often enables closed-form results (in~\cite{cui2022channel},~\cite{kosasih2025spatialfrequencies}, and~\cite{lu2022communicating}, among others). However, the idealized point Rx model limits the results' applicability to practical NF scenarios with a realistic non-point Rx array.

The second group of NF beamfocusing studies models the Rx as an array but often employs \emph{non-analytical} combining techniques, such as analog equal-gain combining (EGC) and digital maximum-ratio combining (MRC)~\cite{shi2025doublesided,cui2023nearfieldrainbow}. While offering practical insights, these approaches hinder analytical tractability in follow-up mathematical work (e.g., multi-cell NF interference modeling and coverage optimization), as neither MRC nor EGC features a closed-form analytical structure.

To the best of the authors' knowledge, only a few studies have attempted to analytically characterize NF Rx combiners for beamfocusing. Among those, Dardari established asymptotic limits for NF and far-field (FF) links using ideal continuous arrays in~\cite{dardari2019communicating}. Recently, Chen et al.~\cite{chen2026receiveraware} investigated an \emph{Rx-aware beamfocusing} wavefront for planar arrays, while the authors in~\cite{mizmizi2026focusingconfocusing} analytically studied a similar con-focusing approach for uniform linear arrays (ULAs). However, analytically combining received NF signals across an extended Rx aperture to maximize the signal-to-noise ratio (SNR), and characterizing the impact of Rx array size and combining strategy on achievable SNR limits, \emph{remain insufficiently addressed}.

To address this gap, we first develop two \emph{analytical} Rx combiners in Section~\ref{sec:analytical_characterization} for the line-of-sight (LoS) NF channel between aligned and misaligned Tx and Rx ULAs, where the Tx performs NF beamfocusing. The first, \emph{analog} NF combiner (ANF--AC), employs Rx-side phase compensation and phase-flip corrections (PFCs). The second, \emph{digital} NF combiner ANF--DC, additionally exploits amplitude variations through MRC. We utilize the \emph{resulting analytical structures} of these combiners in Section~\ref{sec:optimal_receiver_aperture} to: (i)~derive a \emph{closed-form approximation for the optimal Tx--Rx array size trade-off} under analog Rx combining; and (ii)~characterize the \emph{SNR gap between analog and digital NF combiners} under Tx-side NF beamfocusing. Section~\ref{sec:results} presents the main numerical results, confirming the near-optimal performance of both proposed analytical combiners. Section~\ref{sec:conclusion} concludes the paper.

%%%%%%%%%%%%%%%%%%%%%%%%%%%%%%%%%%%%%%%%%%%%%%%%%%%%%%%%%%%%%%%%%%%%%%%%%%%%%%%%%%%%%%%%%%%%%%%%%%%%%%%%%%%%%%%%%%%%%%%%%%%%%%%%%%%%%%%
%%%%%%%%%%%%%%%%%%%%%%%%%%%%%%%%%%%%        System Model        %%%%%%%%%%%%%%%%%%%%%%%%%%%%%%%%%%%%%%%%%%%%%%%%%%%%%%%%%%%%%%%%%%%%%%%
%%%%%%%%%%%%%%%%%%%%%%%%%%%%%%%%%%%%%%%%%%%%%%%%%%%%%%%%%%%%%%%%%%%%%%%%%%%%%%%%%%%%%%%%%%%%%%%%%%%%%%%%%%%%%%%%%%%%%%%%%%%%%%%%%%%%%%%
\section{System Model}
\label{sec:system_model}
\begin{figure}[t]
    \centering
    \includegraphics[width=0.98\columnwidth]
    {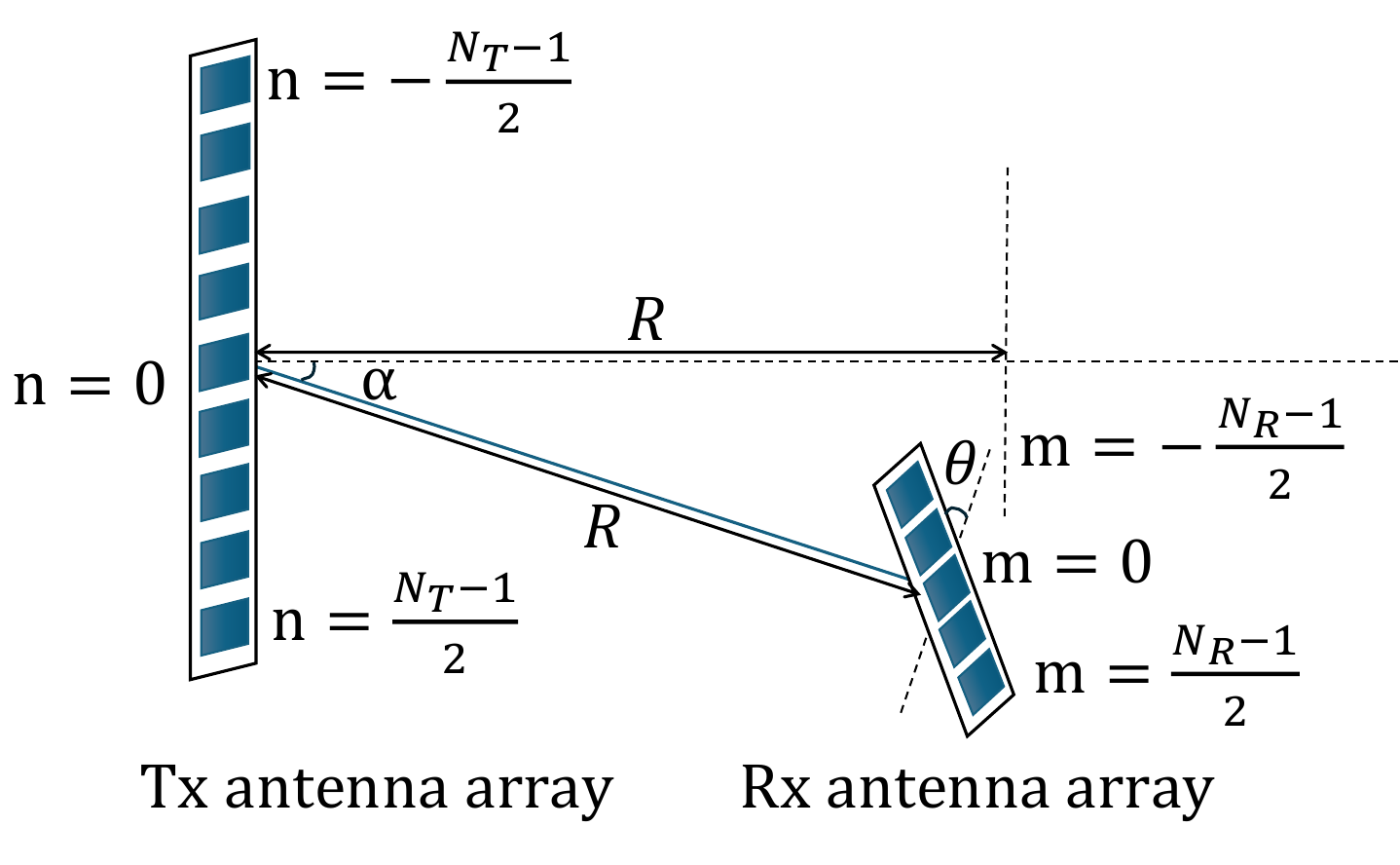}
    \vspace{-2mm}
    \caption{System model and the antenna array geometry. Near-field link between two misaligned ULAs of $N_{\text{T}}$ and $N_{\text{R}}$ elements. Tx applies near-field beamfocusing toward the center of the Rx. \emph{Rx explores different signal combining strategies.}}
    \label{fig:system_model}
    %\vspace{-1mm}
\end{figure}

We model a point-to-point LoS free-space communication system operating in the radiative near-field region, as in Fig.~\ref{fig:system_model}. The Tx and Rx employ ULAs with $N_{\text{T}}$ and $N_{\text{R}}$ antenna elements, respectively. To account for practical misalignment, we introduce the geometric angles $\alpha$ and $\theta$, representing the respective Tx and Rx ULA rotations (we consider in-plane misalignment, with both ULAs lying in the same plane). The inter-element spacing is $\lambda/2$, where $\lambda$ is the carrier wavelength. We neglect secondary-order effects due to mutual coupling or polarization mismatch. For clarity, we assume odd-sized Tx and Rx arrays with centered indices.\footnote{Without loss of generality, the analysis extends to even-sized arrays using half-integer indices, complicating the formulas but keeping the key results.} The distance between the Tx and Rx array centers is $R$ meters.

\emph{Notation:} The index set of an $N$-element array is defined as
$\mathcal{I}_N =
\left\{
-\frac{N-1}{2},\ldots,\frac{N-1}{2}
\right\},
\,\,N\in\{N_{\text{T}},N_{\text{R}}\}$.
The $m$-th Rx and $n$-th Tx element positions, relative to their respective array centers and measured along their array axes, are
$p_m=m\frac{\lambda}{2}$ and $p_n=n\frac{\lambda}{2}$, respectively, where $m\in\mathcal{I}_{N_{\text{R}}}$ and $n\in\mathcal{I}_{N_{\text{T}}}$.

%%%%%%%%%%%%%%%%%%%%
\subsection{Near-Field Channel Model}
\label{subsec:channel_model}
\vspace{-0.5mm}
The channel matrix $\mathbf{H}\in\mathbb{C}^{N_{\text{R}}\times N_{\text{T}}}$ describes propagation between the ULAs. Its $(m,n)$-th element, $h_{m,n}$, represents the complex channel coefficient between Tx element $n$ and Rx element $m$. Under the NF spherical-wave model, it is~\cite{liu2023nearfield}:
\vspace{-2mm}
\begin{equation}
    h_{m,n}
    =
    \frac{\lambda}{4\pi d_{m,n}}
    \exp\left(-j\frac{2\pi}{\lambda}d_{m,n}\right),
    \label{eq:channel_coefficient}
    \vspace{-1mm}
\end{equation}
where $d_{m,n}$ denotes the distance between the corresponding Tx and Rx elements. Based on the array geometry in Fig.~\ref{fig:system_model},
\vspace{-2mm}
\begin{align}
d_{m,n}
= \Big[&
\big(R\sin\alpha
+p_m\cos(\alpha-\theta)-p_n\big)^2
\nonumber\\
&\quad+
\big(R\cos\alpha
-p_m\sin(\alpha-\theta)\big)^2
\Big]^{1/2}.
\label{eq:element_distance}
\vspace{-1mm}
\end{align}
This NF model captures distance-dependent amplitude attenuation and propagation phase for each Tx--Rx element path~\cite{liu2023nearfield}.

%%%%%%%%%%%%%%%%%%%%
\subsection{Near-Field Beamfocusing}
\label{subsec:transmit_precoder}
\vspace{-0.5mm}
We employ a phase-only precoder that focuses the transmitted signal toward the center of the Rx array ($m=0$ in Fig.~\ref{fig:system_model}). The precoder coefficients are therefore given by
\vspace{-1.5mm}
\begin{equation}
    [\mathbf{w}_t]_n
    =
    \frac{1}{\sqrt{N_{\text{T}}}}
    \exp\left(j\frac{2\pi}{\lambda}d_{0,n}\right),
    \quad n\in\mathcal{I}_{N_{\text{T}}},
    \label{eq:tx_focusing_precoder}
    \vspace{-0.5mm}
\end{equation}
where $d_{0,n}$ is the distance between Tx element $n$ and the Rx array center, accounting for possible Tx--Rx misalignment as illustrated in Fig.~\ref{fig:system_model}. The precoder has a unit norm, $\|\mathbf{w}_t\|_2=1$.

%%%%%%%%%%%%%%%%%%%%
\vspace{-1mm}
\subsection{Received Signal and SNR}
\label{subsec:received_signal_snr}
\vspace{-1mm}
The received signal after receive combining is
\begin{equation}
    z
    =
    \sqrt{P_{\text{T}}}\mathbf{w}_r^H\mathbf{H}\mathbf{w}_t s
    +\mathbf{w}_r^H\mathbf{n},
    \label{eq:combined_received_signal}
\end{equation}
where $P_{\text{T}}$ denotes the total Tx power spectral density (p.s.d.), $s$ is a unit-power information symbol, and $\mathbf{n}\sim\mathcal{CN}(\mathbf{0},N_0\mathbf{I}_{N_{\text{R}}})$ is the Rx noise vector, where $N_0$ is the per-branch noise p.s.d.

The unit-norm combiner $\mathbf{w}_r$ determines the combination of signals across the Rx. The resulting SNR with $\|\mathbf{w}_r\|_2=1$ is
\vspace{-2mm}
\begin{equation}
    \mathrm{SNR}
    =
    \frac{P_{\text{T}}}{N_0}
    \left|\mathbf{w}_r^H\mathbf{H}\mathbf{w}_t\right|^2.
    \label{eq:received_snr}
    \vspace{-1mm}
\end{equation}

The constraints on $\mathbf{w}_r$ depend on the Rx architecture. We compare two main implementations: (i)~fully digital, allowing independent amplitude and phase control, and (ii)~phase-only analog combining, satisfying $|[\mathbf{w}_r]_m| = \frac{1}{\sqrt{N_{\text{R}}}}, \,\, m\in\mathcal{I}_{N_{\text{R}}}$.

%%%%%%%%%%%%%%%%%%%%%%%%%%%%%%%%%%%%%%%%%%%%%%%%%%%%%%%%%%%%%%%%%%%%%%%%%%%%%%%%%%%%%%%%%%%%%%%%%%%%%%%%%%%%%%%%%%%%%%%%%%%%%%%%%%%%%%%
%%%%%%%%%%%%%%%%%%%%%%%%%%%%%%%%%%%%        Section III. Combiners        %%%%%%%%%%%%%%%%%%%%%%%%%%%%%%%%%%%%%%%%%%%%%%%%%%%%%%%%%%%%%
%%%%%%%%%%%%%%%%%%%%%%%%%%%%%%%%%%%%%%%%%%%%%%%%%%%%%%%%%%%%%%%%%%%%%%%%%%%%%%%%%%%%%%%%%%%%%%%%%%%%%%%%%%%%%%%%%%%%%%%%%%%%%%%%%%%%%%%
\section{Analytical Near-Field Combining Strategies}
\label{sec:analytical_characterization}
We first review two conventional, \emph{non-analytical} yet theoretically optimal analog and digital combiners as benchmarks. We then propose two \emph{analytical} NF beamfocusing-specific schemes featuring both \emph{comparable performance} (Sec.~\ref{sec:results}) and \emph{analytical tractability} for network analysis and optimization.

With the Tx precoder $\mathbf{w}_t$ \eqref{eq:tx_focusing_precoder}, the effective channel vector~is
\vspace{-2mm}
\begin{equation}
    \mathbf{g}=\mathbf{H}\mathbf{w}_t \in \mathbb{C}^{N_{\text{R}}}.
    \label{eq:effective_channel}
    \vspace{-1mm}
\end{equation}

With a fully digital Rx, MRC is SNR-maximizing under the unit-norm constraint~\cite{proakis2001digital}~and~is given by
\vspace{-2mm}
\begin{equation}
    \mathbf{w}_r^{\mathrm{MRC}}
    =\frac{\mathbf{g}}{\|\mathbf{g}\|_2}.
    \label{eq:mrc_combiner}
    \vspace{-1mm}
\end{equation}
With an analog Rx with a constant-modulus constraint, EGC is the corresponding SNR-maximizing method\cite{proakis2001digital}:
\vspace{-1mm}
\begin{equation}
    [\mathbf{w}_r^{\mathrm{EGC}}]_m
    =\frac{1}{\sqrt{N_{\text{R}}}}
    \exp\left(j\arg(g_m)\right),
    \quad m\in\mathcal{I}_{N_{\text{R}}},
    \label{eq:egc_combiner}
    \vspace{-0.5mm}
\end{equation}
where $g_m$ denotes the $m$-th element of $\mathbf{g} = [g_1,\ldots,g_{N_{\text{R}}}]^{T}$.

MRC and EGC serve as reference benchmarks for the proposed NF digital and analog combining schemes, respectively.

%%%%%%%%%%%%%%%%%%%%
\subsection{Closed-Form Approximation of the Effective NF Channel}
\label{subsec:closed_form_effective_channel}
To construct analytical NF Rx combiners, we first derive a closed-form approximation of the effective NF channel for the considered Tx--Rx setup. This expression is then utilized to derive the proposed combiners in Section~\ref{subsec:proposed_combiners}, with the accuracy of this approximation cross-verified in Section~\ref{sec:results}.

First, recalling that $m\in\mathcal{I}_{N_{\text{R}}}$ (see Sec.~\ref{sec:system_model}), the $m$-th element of the effective channel vector in
\eqref{eq:effective_channel} can be computed as
\vspace{-0.5mm}
\begin{equation}
\begin{aligned}
\hspace{-1mm}g_m\hspace{-0.5mm}=\hspace{-0.5mm}\frac{1}{\sqrt{N_{\text{T}}}}\hspace{-0.5mm}
\sum_{n\in\mathcal{I}_{N_{\text{T}}}}
\hspace{-0.5mm}\frac{\lambda}{4\pi d_{m,n}} \hspace{-0.5mm}\hspace{-0.5mm}
\exp\left(
-j\frac{2\pi}{\lambda}
(d_{m,n}-d_{0,n})
\hspace{-0.5mm}\right).\hspace{-0.5mm}
\end{aligned}
\label{eq:effective_channel_element}
\end{equation}

We then approximate \(d_{m,n}-d_{0,n}\) via a second-order Taylor polynomial to simplify the phase term. From \eqref{eq:element_distance}, the distance between the $n$-th Tx and $m$-th Rx array elements is
\vspace{-1mm}
\begin{align}
d_{m,n}
=R\Bigg[
1
&+\frac{2(p_m\sin\theta-p_n\sin\alpha)}{R}
\nonumber\\
&+\frac{
p_m^2-2p_mp_n\cos(\alpha-\theta)+p_n^2
}{R^2}
\Bigg]^{1/2}.
\label{eq:element_distance_factored}
\vspace{-2mm}
\end{align}
Using the approximation
$\sqrt{1+\varepsilon}\approx
1+\varepsilon/2-\varepsilon^2/8$, we get
\vspace{-1mm}
\begin{align}
\hspace{-2mm}d_{m,n}\hspace{-0.5mm}\approx \hspace{-0.5mm} R
\hspace{-0.5mm}+\hspace{-0.5mm}p_m\sin\theta \hspace{-0.5mm} - \hspace{-0.5mm} p_n\sin\alpha \hspace{-0.5mm} + \hspace{-0.5mm} \frac{(p_m\cos\theta-p_n\cos\alpha)^2}{2R},
\label{eq:element_distance_approximation}
\vspace{-3mm}
\end{align}
where terms of order $1/R^2$ from the
$\varepsilon^2/8$-term are neglected.

Finally, using the uniform spherical-wave model~\cite{liu2023nearfield}, we approximate \(d_{m,n}\) in \eqref{eq:effective_channel_element} with \(R\) in the amplitude term, making it independent of \(m\). Substituting the distance from \eqref{eq:element_distance_approximation} into the phase term of \eqref{eq:effective_channel_element}, along with the element positions $p_m$ and $p_n$, yields the effective channel approximation $g_m$ as
\vspace{-1mm}
\begin{equation}
\begin{aligned}
g_m
&\approx
\frac{\lambda}{4\pi R\sqrt{N_{\text{T}}}}
e^{-j\phi_m}
\sum_{n\in\mathcal{I}_{N_{\text{T}}}}e^{j2mn\beta}\\
&=
\frac{\lambda}{4\pi R\sqrt{N_{\text{T}}}}
e^{-j\phi_m}
\frac{\sin(N_{\text{T}} m\beta)}{\sin(m\beta)},
\end{aligned}
\label{eq:effective_channel_closed_form}
\vspace{-1mm}
\end{equation}
where
\vspace{-1mm}
\begin{equation}
\phi_m \triangleq
\pi m\sin\theta
+\frac{\pi m^2\lambda\cos^2\theta}{4R},\quad{} \beta \triangleq \frac{\pi\lambda\cos\alpha\cos\theta}{4R}.
\label{eq:rx_phase}
\vspace{-1mm}
\end{equation}

Note that the sine ratio in \eqref{eq:effective_channel_closed_form} is interpreted by continuity whenever its denominator vanishes; its value at \(m=0\) is \(N_{\text{T}}\).

%%%%%%%%%%%%%%%%%%%%
\subsection{Proposed Analytical Near-Field Combiners}
\label{subsec:proposed_combiners}
Using the effective NF channel approximation in \eqref{eq:effective_channel_closed_form}, we propose two geometry-based Rx combiners: \emph{ANF--AC} and~\emph{ANF--DC}. Both exploit $g_m$'s phase structure, while digital ANF--DC also captures its amplitude variations across the Rx.

\subsubsection{Analytical Near-Field Analog Combiner (ANF--AC)}
\label{subsubsec:anf_ac}
The EGC combiner in \eqref{eq:egc_combiner} compensates each effective channel coefficient's phase using constant-modulus weights. Based on the closed-form approximation in \eqref{eq:effective_channel_closed_form}, we propose the analytical NF analog combiner (ANF--AC), with coefficients
\begin{equation}
\boxed{[\mathbf{w}_r^{\mathrm{ANF\text{-}AC}}]_m
=
\frac{1}{\sqrt{N_{\text{R}}}}
\exp\left(
-j\phi_m+j\pi\delta_m^{\mathrm{PFC}}
\right),}
\label{eq:anf_ac_combiner}
\end{equation}
for $m\in\mathcal{I}_{N_{\text{R}}}$, where $\phi_m$ denotes the geometric phase term and $\delta_m^{\mathrm{PFC}}$ denotes the phase-flip correction (PFC), defined as
\begin{equation}
\delta_m^{\mathrm{PFC}}
=
\left\lfloor\frac{N_{\text{T}} m\beta}{\pi}\right\rfloor
-
\left\lfloor\frac{m\beta}{\pi}\right\rfloor.
\label{eq:pfc_definition}
\end{equation}

The ANF--AC structure comprises two phase-only compensation components: (i)~the focusing-back phase $\phi_m$, directing the received wavefront toward the Tx array center, and (ii)~the proposed PFCs, compensating $\pi$-phase reversals caused by sign changes in the sine-ratio term in \eqref{eq:effective_channel_closed_form}.

%%%%%%%%%%%%%%%%%%%%
\subsubsection{Analytical Near-Field Digital Combiner (ANF--DC)}
\label{subsubsec:anf_dc}
%For fully digital reception, the unit-norm MRC vector in \eqref{eq:mrc_combiner} is SNR-optimal under a unit-norm constraint~\cite{proakis2001digital}.
Substituting the channel approximation in \eqref{eq:effective_channel_closed_form} into \eqref{eq:mrc_combiner} yields ANF--DC. Like ANF--AC, it applies geometric phase compensation toward the Tx-array center and PFCs for Dirichlet-kernel sign reversals~\cite{katznelson2004harmonic}. Further, the digital ANF--DC exploits channel magnitudes across the Rx, assigning larger weights to elements with stronger effective channels.

The $m$-th coefficient of the ANF--DC is hence derived as
\begin{equation}
\hspace{-2mm}\boxed{[\widetilde{\mathbf{w}}_r^{\mathrm{ANF\text{-}DC}}]_m
\hspace{-0.5mm}=\hspace{-0.5mm}
\left|\frac{\sin(N_{\text{T}} m\beta)}{\sin(m\beta)}\right|
\exp\left(
-j\phi_m \hspace{-0.5mm}+\hspace{-0.5mm} j\pi\delta_m^{\mathrm{PFC}}
\right),\hspace{-0.5mm}}
\label{eq:anf_dc_combiner_unnormalized}
\end{equation}
while the normalized ANF--DC combining vector becomes
\vspace{-1mm}
\begin{equation}
\mathbf{w}_r^{\mathrm{ANF\text{-}DC}}
=
\frac{\widetilde{\mathbf{w}}_r^{\mathrm{ANF\text{-}DC}}}
{\left\|\widetilde{\mathbf{w}}_r^{\mathrm{ANF\text{-}DC}}\right\|_2}.
\label{eq:anf_dc_combiner}
\end{equation}

%%%%%%%%%%%%%%%%%%%%%%%%%%%%%%%%%%%%%%%%%%%%%%%%%%%%%%%%%%%%%%%%%%%%%%%%%%%%%%%%%%%%%%%%%%%%%%%%%%%%%%%%%%%%%%%%%%%%%%%%%%%%%%%%%%%%%%%
%%%%%%%%%%%%%%%%%%%%%%%%%%%%%%%%%%%%        Section IV. Analysis        %%%%%%%%%%%%%%%%%%%%%%%%%%%%%%%%%%%%%%%%%%%%%%%%%%%%%%%%%%%%%%%
%%%%%%%%%%%%%%%%%%%%%%%%%%%%%%%%%%%%%%%%%%%%%%%%%%%%%%%%%%%%%%%%%%%%%%%%%%%%%%%%%%%%%%%%%%%%%%%%%%%%%%%%%%%%%%%%%%%%%%%%%%%%%%%%%%%%%%%
\section{Optimal Receiver Size for NF Beamfocusing}
\label{sec:optimal_receiver_aperture}
\vspace{-0.8mm}
In this section, we evaluate ANF--AC and ANF--DC, then leverage their \emph{analytical} properties to derive a theoretical result and two engineering insights for future NF wireless systems.
\vspace{-2mm}
\subsection{Combining Gains of the ANF--AC and ANF--DC}
\label{subsec:combining_gain_characterization}
\vspace{-1mm}
To evaluate the end-to-end performance of our NF system with the proposed combiners, we recall the $\mathrm{SNR}$ in \eqref{eq:received_snr} that can be expressed in terms of the effective channel vector \(\mathbf{g}\) as
\vspace{-2mm}
\begin{align}
\mathrm{SNR}=\frac{P_{\text{T}}}{N_0}
\underbrace{\left|\mathbf{w}_r^H\mathbf{g}\right|^2}_{G_{\mathrm{Rx}}}.
\vspace{-0mm}
\end{align}
With fixed propagation geometry and Tx precoding, the effective channel is unchanged, making the SNR proportional to the combining gain \(G_{\mathrm{Rx}}\). Hence, to evaluate the SNR at the Rx, we derive the combining gains of ANF--AC and ANF--DC.

To obtain the combining gain with ANF--AC, $G_{\mathrm{ANF\text{-}AC}}$, we substitute \eqref{eq:rx_phase} and \eqref{eq:pfc_definition} into \eqref{eq:anf_ac_combiner}. We then sum the result over all the Rx antenna array elements ($N_{\text{R}}$ in total), leading to
\begin{equation}
\begin{aligned}
G_{\mathrm{ANF\text{-}AC}}
&\triangleq
\left|
(\mathbf{w}_r^{\mathrm{ANF\text{-}AC}})^H
\mathbf{H}\mathbf{w}_t
\right|^2\\
&=
\frac{\lambda^2}{16\pi^2 R^2N_{\text{T}}N_{\text{R}}}
\left(
\sum_{m\in\mathcal{I}_{N_{\text{R}}}}
\left|
\frac{\sin(N_{\text{T}}\beta m)}{\sin(\beta m)}
\right|
\right)^2.
\end{aligned}
\label{eq:receiver_gain_analog_exact}
\end{equation}

We can further approximate the resulting sum by an integral using a Riemann-sum approximation~\cite{stewart2008calculus} and then use the small-angle approximation $\sin(\beta m)\approx\beta m$, leading to 
\begin{equation}
\begin{aligned}
\hspace{-2mm}G_{\mathrm{ANF\text{-}AC}}
\hspace{-0.5mm}&\approx\hspace{-0.5mm}
\frac{\lambda^2}{16\pi^2 R^2}
\frac{4}{\beta^2N_{\text{T}}N_{\text{R}}}
\left(\hspace{-0.5mm}
\int_0^{N_{\text{T}}N_{\text{R}}\beta/2}
\hspace{-0.5mm}\frac{|\sin u|}{u}\hspace{-0.5mm}\,du
\hspace{-0.5mm}\right)^2.\hspace{-1.5mm}
\end{aligned}
\label{eq:receiver_gain_analog_approx}
\end{equation}

Similarly, the combining gain with the digital ANF--DC is
\begin{equation}
\begin{aligned}
G_{\mathrm{ANF\text{-}DC}}
&\triangleq
\left|
(\mathbf{w}_r^{\mathrm{ANF\text{-}DC}})^H
\mathbf{H}\mathbf{w}_t
\right|^2\\
&=
\frac{\lambda^2}{16\pi^2 R^2N_{\text{T}}}
\sum_{m\in\mathcal{I}_{N_{\text{R}}}}
\left|
\frac{\sin(N_{\text{T}}\beta m)}{\sin(\beta m)}
\right|^2\\
&\approx
\frac{\lambda^2}{16\pi^2 R^2}
\frac{2}{N_{\text{T}}\beta}
\int_0^{\beta N_{\text{R}}/2}
\left(
\frac{\sin(N_{\text{T}} u)}{\sin u}
\right)^2du.
\end{aligned}
\label{eq:receiver_gain_digital_approx}
\end{equation}

%%%%%%%%%%%%%%%%%%%%
\subsection{Which Size of Analog Rx Array is Best for Beamfocusing?}
\label{subsec:optimal_aperture_derivation}
We now leverage the analytical property of the ANF--AC and ANF--DC combiners to obtain new results. We first calculate the \emph{optimal size of the Rx array}, $N_{\text{R}}$, that maximizes the combining gain of the ANF--AC, $G_{\mathrm{ANF\text{-}AC}}$, and consequently (see above) \emph{maximizes the SNR at the Rx}. For this, we define
\vspace{-0.5mm}
\begin{equation}
J(y)\triangleq\int_0^y\frac{|\sin u|}{u}\,du,
\qquad
y\triangleq\frac{N_{\text{T}}N_{\text{R}}\beta}{2}.
\label{eq:receiver_optimization_functions}
\vspace{-0.5mm}
\end{equation}
Then, substituting the definition of $y$ in \eqref{eq:receiver_optimization_functions} into \eqref{eq:receiver_gain_analog_approx} leads~to
\vspace{-0.5mm}
\begin{equation}
G_{\mathrm{ANF\text{-}AC}}
\approx
\frac{\lambda^2}{16\pi^2 R^2}
\frac{2}{\beta}
\frac{J^2(y)}{y}.
\label{eq:anf_ac_gain_y}
\vspace{-0.5mm}
\end{equation}

Since the prefactor in \eqref{eq:anf_ac_gain_y} is independent of $y$, maximizing $G_{\mathrm{ANF\text{-}AC}}$ is equivalent to maximizing $J^2(y)/y$. Differentiating with respect to $y$ and setting the derivative to zero gives
\vspace{-2mm}
\begin{equation}
2yJ'(y)=J(y).
\label{eq:receiver_optimality_condition}
\vspace{-0.5mm}
\end{equation}
The first positive stationary point is obtained numerically as
\vspace{-2mm}
\begin{equation}
y^\star\approx 2.15.
\label{eq:optimal_y}
\vspace{-1mm}
\end{equation}

Finally, substituting this value into \eqref{eq:receiver_optimization_functions} gives the \emph{optimal transmit--receive antenna-count product} maximizing the SNR for analog reception with Tx beamfocusing toward the Rx:
\begin{equation}
\boxed{(N_{\text{T}}N_{\text{R}})^\star
\approx
\frac{5.48\,R}
{\lambda\cos\alpha\cos\theta}.}
\label{eq:optimal_antenna_product}
\end{equation}

\emph{Discussion:} Note that this result applies to both perfectly aligned Tx and Rx ULAs and more realistic scenarios where the Rx lies outside the Tx broadside ($\alpha \neq 0$) and/or is misaligned with the Tx center (rotation, $\theta \neq 0$). Equation \eqref{eq:optimal_antenna_product} also captures the key parameters from Section~\ref{sec:system_model} and Fig.~\ref{fig:system_model}---separation $R$, wavelength $\lambda$, array sizes ($N_{\text{T}}$ and $N_{\text{R}}$), and orientations ($\alpha$ and $\theta$)---in a \emph{simple closed-form expression}.\footnote{To the best of the authors' knowledge, obtaining a similar result in the near field for \emph{non-analytical} analog Rx combiners (such as e.g., a canonical EGC), is possible only numerically. We further discuss the implications and utilization of \eqref{eq:optimal_antenna_product} in the analysis below and in the numerical results.}

%%%%%%%%%%%%%%%%%%%%
\subsection{Tx and Rx Size Trade-off for Near-field Beamfocusing}
\label{subsec:optimal_beam_capture_level}
\begin{figure}[!t]
    \centering
    \includegraphics[width=1\linewidth]{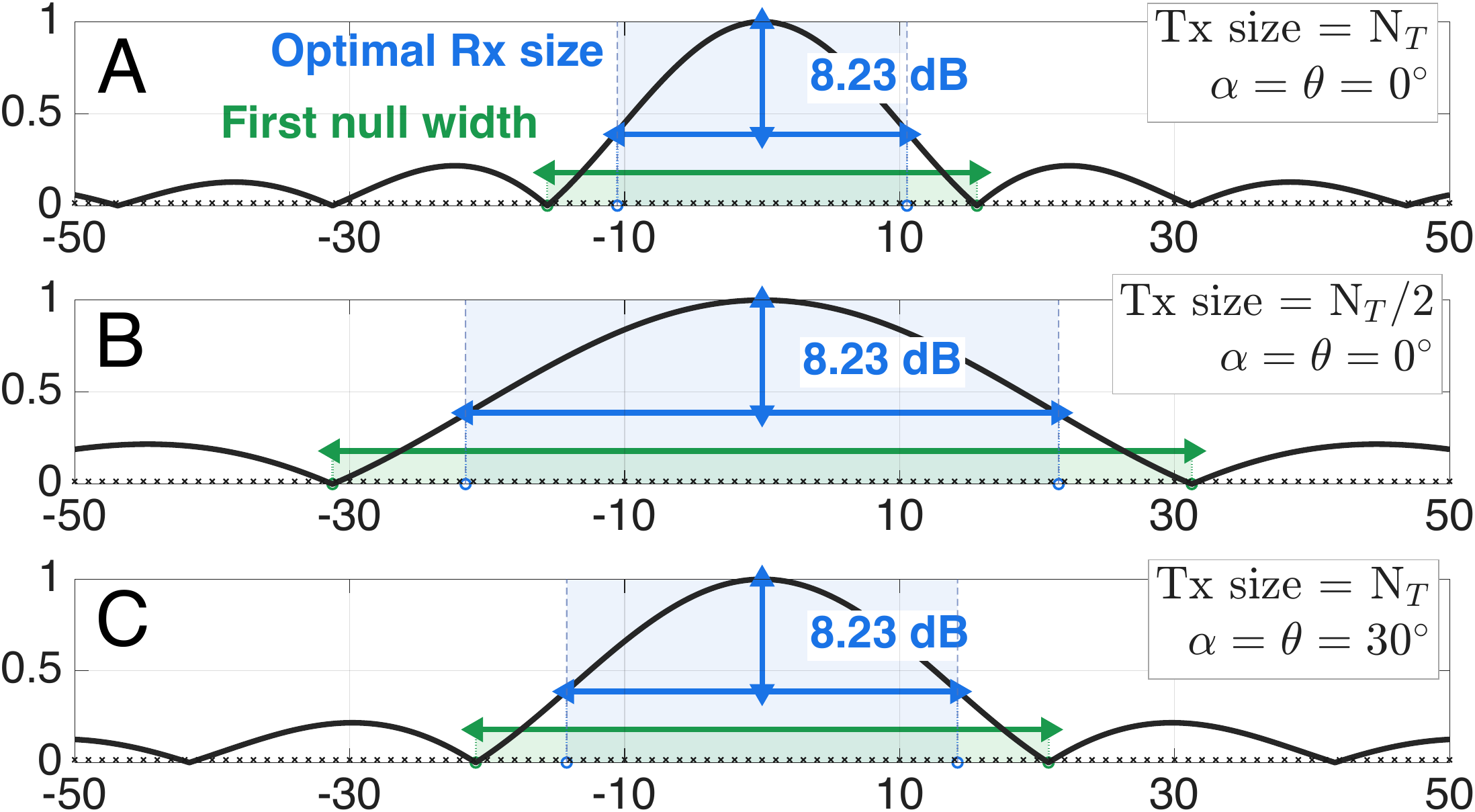}
    \caption{Amplitude profile across the Rx array under near-field beamfocusing. The first-null aperture width (green), $\approx$~$4R / \left( N_{\text{T}} \cos{\alpha} \cos{\theta} \right)$, defines the main lobe. The blue width -- the optimal $N_{\text{R}}$ value in \eqref{eq:optimal_antenna_product} to maximize the SNR.}
    \label{fig:amplitude_profile}
    \vspace{-1mm}
\end{figure}

Our result in \eqref{eq:optimal_antenna_product} demonstrates that increasing the Rx array $N_{\text{R}}$ is not always desirable in the NF LoS scenario with Tx beamfocusing. The Tx--Rx array size trade-off ($N_{\text{T}}$ and $N_{\text{R}}$) for maximizing SNR follows from a key NF beamfocusing property: at a fixed distance $R$, larger Tx arrays produce tighter focal spots around the Rx center than smaller ones~\cite{zhang2022beam}. Fig.~\ref{fig:amplitude_profile} illustrates this effect, where Fig.~\ref{fig:amplitude_profile}(b) uses half the Tx array size from Fig.~\ref{fig:amplitude_profile}a, resulting in a wider focal spot at the Rx.

Further, a phase-only analog Rx combiner ANF--AC adjusts signal phases but not amplitudes or powers before summing across the $N_{\text{R}}$ antennas. Thus, when the Tx focal spot is larger than the Rx aperture, elements outside it have low per-element SNR and contribute mainly noise rather than useful signal. Hence, a smaller Rx array size $N_{\text{R}}$ can be sufficient and even preferable for receiving beams focused by a larger Tx size $N_{\text{T}}$, confirming their inverse dependency in \eqref{eq:optimal_antenna_product}.

A focused beam from a large Tx array features a focal spot longer along the broadside direction than sideways~\cite{lu2024tutorial}. Thus, rotating either ULA in Fig.~\ref{fig:system_model} spreads the Tx beam across more Rx elements: compare Fig.~\ref{fig:amplitude_profile}a with Fig.~\ref{fig:amplitude_profile}(c). This explains the inverse dependence of $N_{\text{T}} N_{\text{R}}$ on $\cos{\alpha}\cos{\theta}$ term in \eqref{eq:optimal_antenna_product}.

\begin{figure*}[!t]
    \centering
    \begin{subfigure}[t]{0.32\textwidth}
        \centering
        \includegraphics[width=\linewidth]{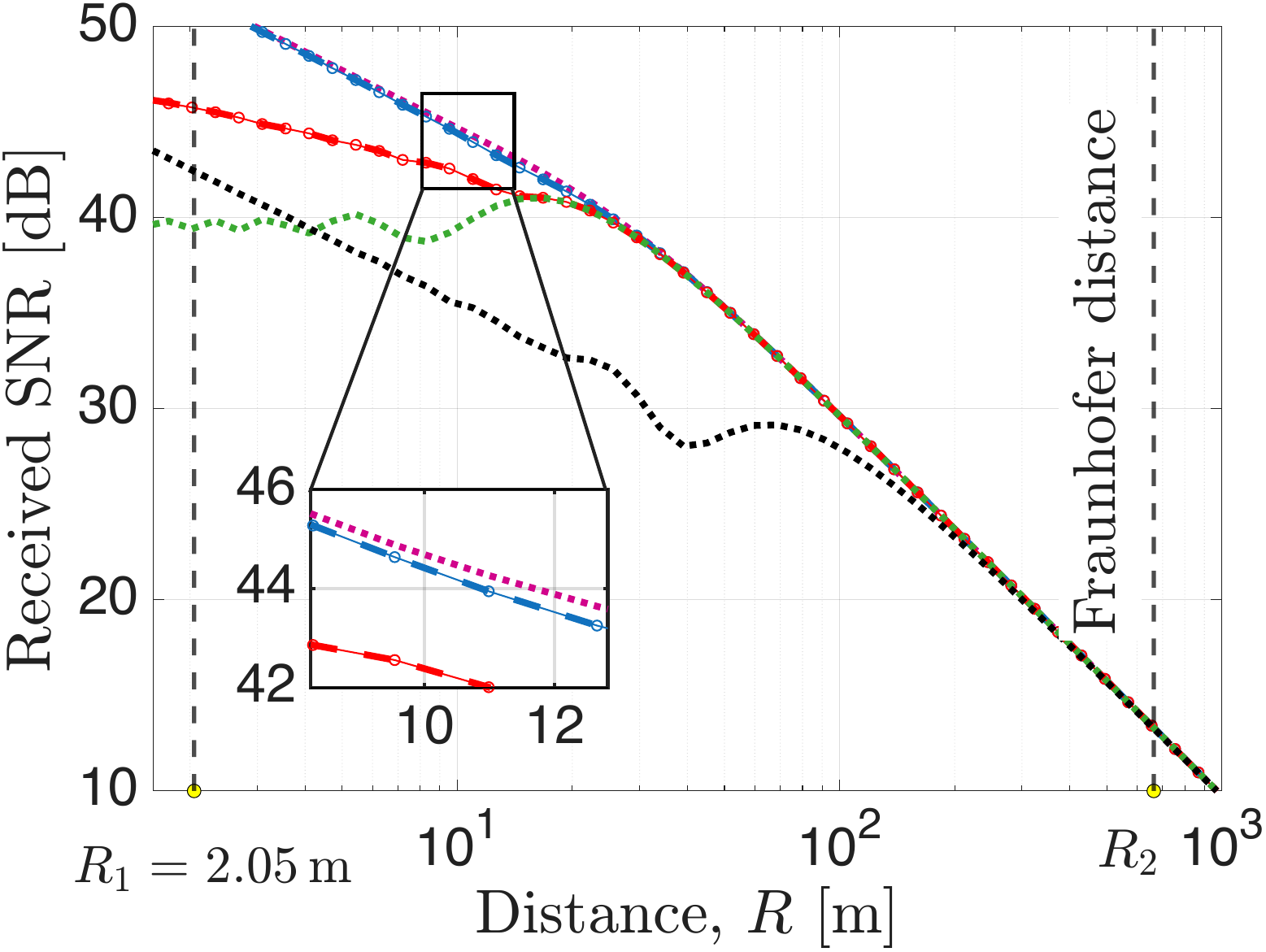}
        \caption{Varying Tx--Rx separation, \emph{legend in (b)}}
        \label{fig:snr_vs_distance}
    \end{subfigure}
    \begin{subfigure}[t]{0.32\textwidth}
        \centering
        \includegraphics[width=\linewidth]{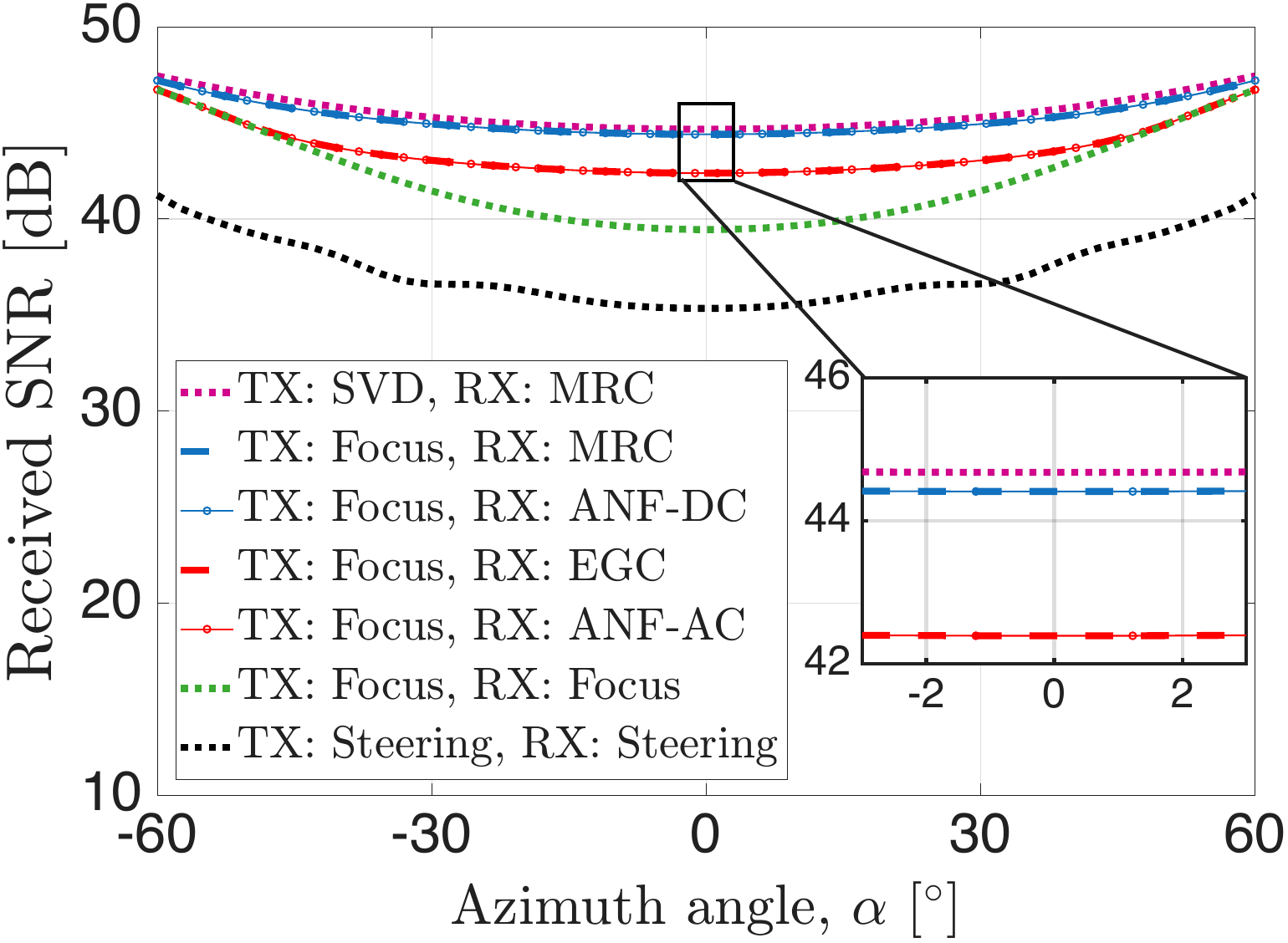}
        \caption{Varying the Tx azimuth angle $\alpha$}
        \label{fig:snr_vs_azimuth}
    \end{subfigure}
    \begin{subfigure}[t]{0.32\textwidth}
        \centering
        \includegraphics[width=\linewidth]{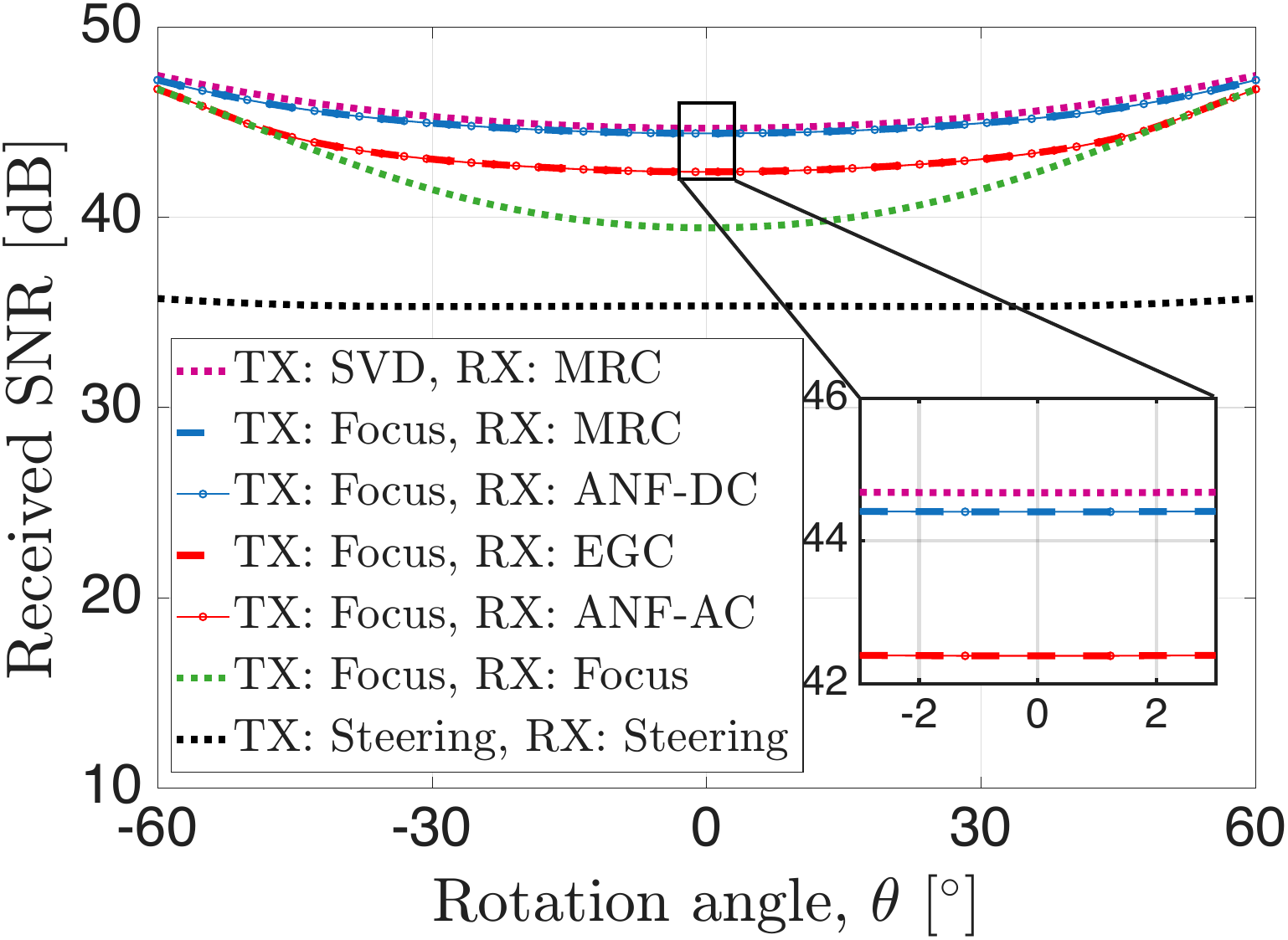}
        \caption{Varying the Rx rotation angle $\theta$}
        \label{fig:snr_vs_rotation}
    \end{subfigure}
    \caption{Comparison of the proposed analytical analog ANF--AC and digital ANF--DC combiners with existing references.}
    \label{fig:downlink_combiner_performance}
    \vspace{-0mm}
\end{figure*}

Finally, for a reasonably large $N_{\text{T}}(>20)$, the approximation $\sin(y^\star/N_{\text{T}})\approx y^\star/N_{\text{T}}$ holds, where $y^\star$ is the optimal value derived in \eqref{eq:optimal_y} for the optimal $N_{\text{R}}$ in \eqref{eq:optimal_antenna_product}. Thus, using the effective channel in \eqref{eq:effective_channel_closed_form}, we obtain the following observation:
\vspace{-1mm}
\begin{align}
&20\log_{10}\left(
\left.\frac{|g_m|}{|g_0|}\right|_{m=N_{\text{R}}^\star/2}
\right)
=
20\log_{10}\left(
\frac{1}{N_{\text{T}}}
\left|
\frac{\sin(y^\star)}
{\sin(y^\star/N_{\text{T}})}
\right|
\right)\notag
\\
&\approx
20\log_{10}\left(
\frac{|\sin(y^\star)|}{y^\star}
\right) \approx \boxed{-8.23~\mathrm{dB}},
\label{eq:optimal_boundary_db}
\vspace{-1mm}
\end{align}
illustrated in Fig.~\ref{fig:amplitude_profile}. Hence, for sufficiently large $N_{\text{T}}$, the optimal focused-beam boundary amplitude is $\approx$$8.23$~dB below the maximum, \emph{independent of other system parameters}.

\begin{figure*}[!t]
    \centering
    \begin{subfigure}{0.48\textwidth}
        \centering
        \includegraphics[width=\linewidth]{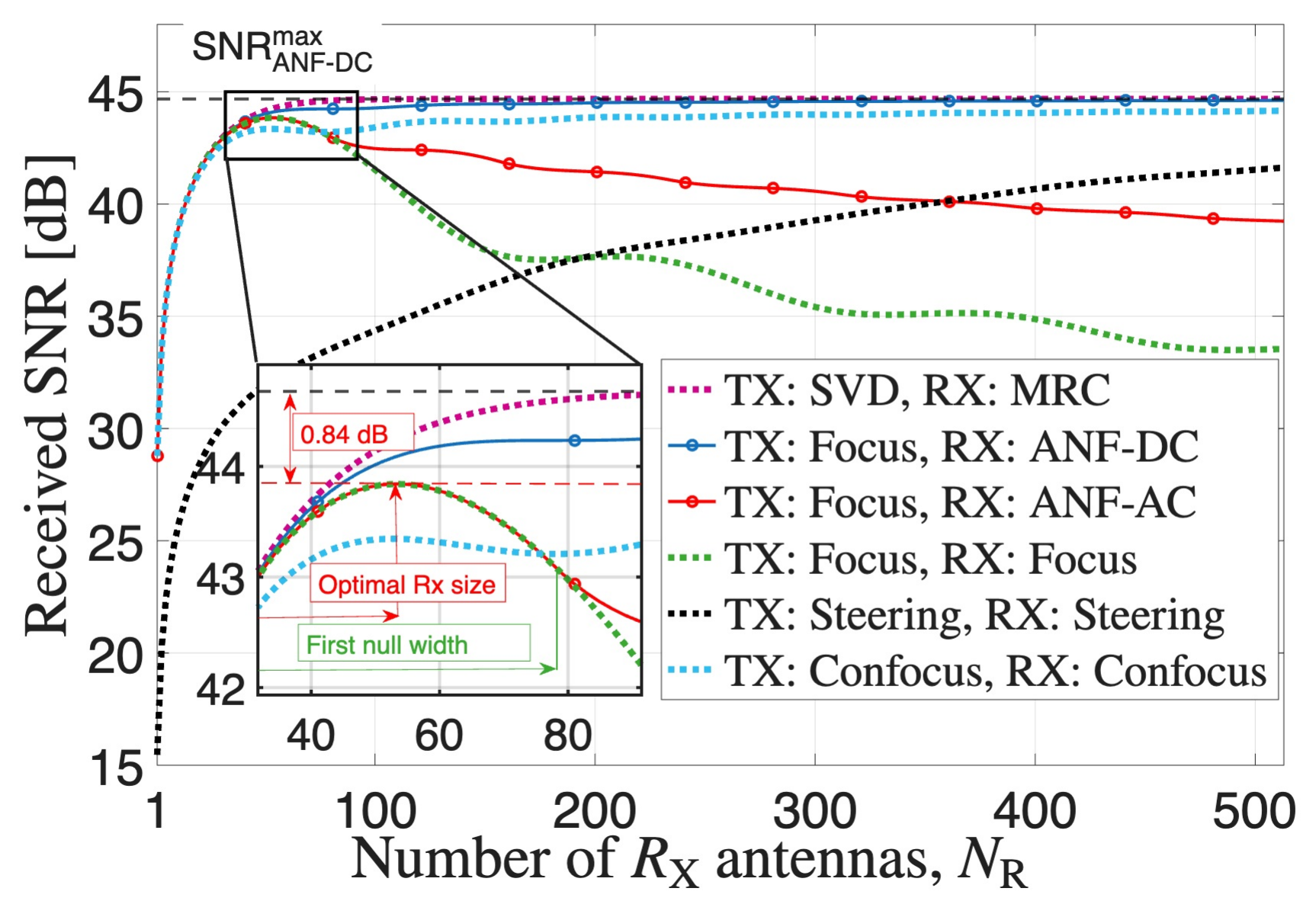}
        \caption{Downlink-type NF scenario, $N_{\text{T}} = 1025$}
        \label{fig:snr_vs_nr}
    \end{subfigure}
    \hfill
    \begin{subfigure}{0.48\textwidth}
        \centering
        \includegraphics[width=\linewidth]{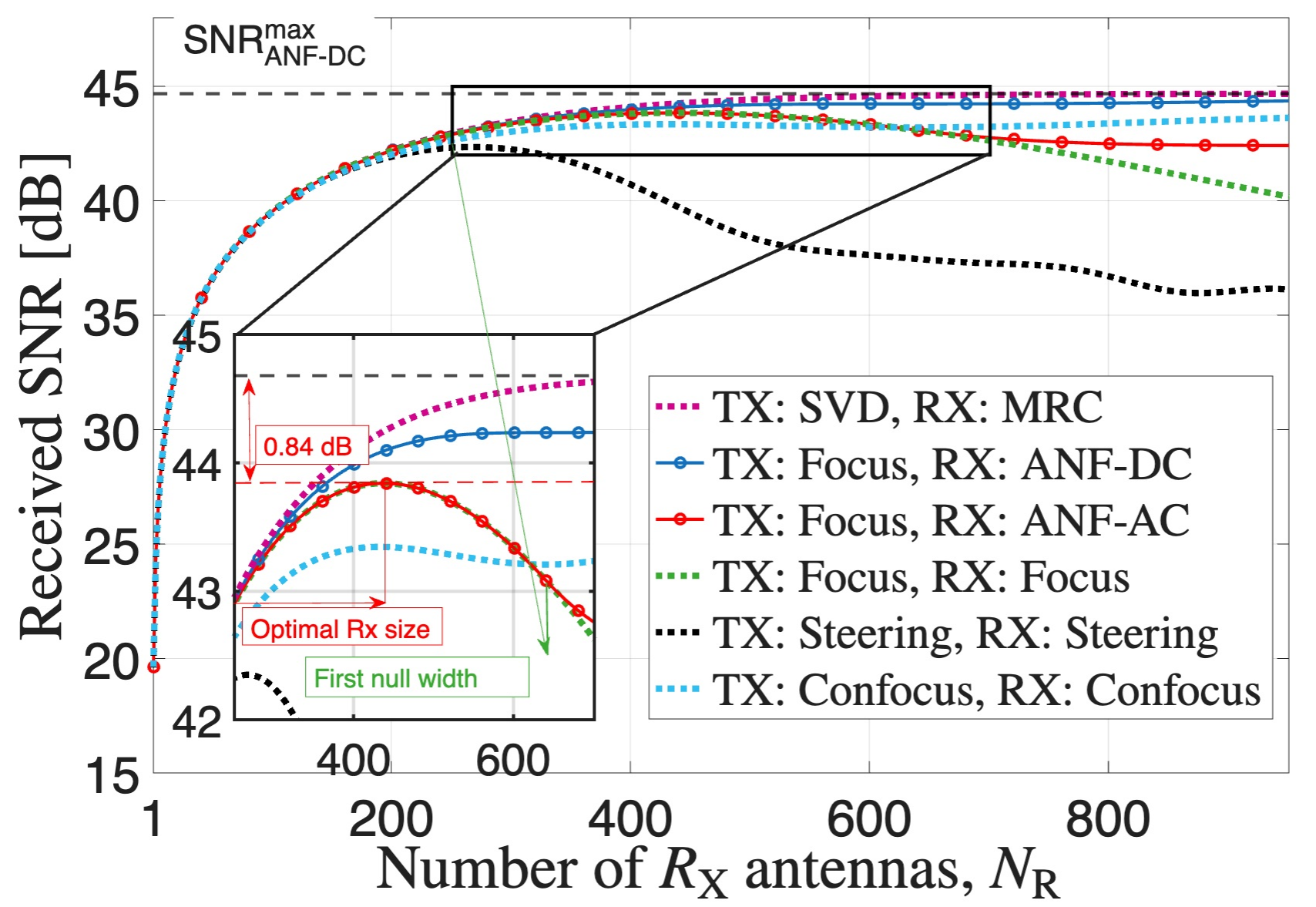}
        \caption{Uplink-type NF scenario, $N_{\text{T}} = 127$}
        \label{fig:snr_vs_nr_uplink}
    \end{subfigure}
    %\vspace{-1mm}
    \caption{Optimal Tx–Rx size trade-off and the corresponding SNR bounds with the proposed NF Rx combiners.}
    \label{fig:snr_vs_nr_comparison}
    \vspace{-2mm}
\end{figure*}

%%%%%%%%%%%%%%%%%%%%
\subsection{SNR Gap Between Fully Digital and Optimized Analog Rx}
\label{subsec:snr_gap_analog_digital}
The presented analytical combiners also reveal an engineering insight: \emph{How much SNR does an optimally designed analog Rx lose relative to a fully digital Rx?} We quantify this loss by comparing the maximum ANF--DC SNR with the SNR by the ANF--AC at its optimal Rx array size. From \eqref{eq:receiver_optimality_condition}, ANF--AC achieves its maximum at $y=y^\star$. For ANF--DC, we use the large-$N_{\text{R}}$ approximation \eqref{eq:receiver_gain_digital_approx}, integrating over $[0,\pi/2]$ (one half-period of the squared Dirichlet kernel) to obtain the large-aperture SNR reference. The resulting maximum SNRs are
\vspace{-2mm}
\begin{align}
&\mathrm{SNR}_{\mathrm{ANF\text{-}AC}}^{\max}
\approx
\frac{P_{\text{T}}}{N_0}
\left(\frac{\lambda}{4\pi R}\right)^2
\frac{2J^2(y^\star)}{\beta y^\star},&
\label{eq:snr_max_analog_digital1}
\vspace{-3mm}
\end{align}
\begin{align}
\label{eq:snr_max_analog_digital2}
&\quad{}\quad{}\mathrm{SNR}_{\mathrm{ANF\text{-}DC}}^{\max}
\triangleq
\lim_{N_{\text{R}}\rightarrow N^{(Max)}_{\text{R}}}
\mathrm{SNR}_{\mathrm{ANF\text{-}DC}}(N_{\text{R}})
\\
&\hspace{-2mm}\approx\hspace{-1mm}
\frac{P_{\text{T}}}{N_0}\hspace{-0.5mm}
\left(\hspace{-0.5mm}\frac{\lambda}{4\pi R}\hspace{-0.5mm}\right)^2
\hspace{-0.5mm}\frac{2}{N_{\text{T}}\beta}
\hspace{-0.5mm}\int_0^{\pi/2}
\hspace{-1mm}\left(\hspace{-0.5mm}
\frac{\sin(N_{\text{T}}u)}{\sin u}
\hspace{-0.5mm}\right)^2\hspace{-1mm}du
=
\frac{P_{\text{T}}}{N_0}
\left(\frac{\lambda}{4\pi R}\right)^2
\frac{\pi}{\beta},\notag
\vspace{-1mm}
\end{align}
where the integral identity holds for integer $N_{\text{T}}$ and $N^{(Max)}_{\text{R}}$ stands for the largest modeled Rx array size $N_{\text{R}}$ when computing the integral. We limit $N^{(Max)}_{\text{R}}$ to $\pi / \beta \approx 40'000$ antennas (a 20-m array at $300$\,GHz and a 200-m array at $30$\,GHz).

Combining \eqref{eq:snr_max_analog_digital1} and \eqref{eq:snr_max_analog_digital2}, we derive the gap between the highest SNR with phase-only analog ANF--AC and the highest SNR with fully digital amplitude-and-phase ANF--DC:
\vspace{-1mm}
\begin{equation}
\begin{aligned}
\Delta_{\mathrm{SNR}}
&\triangleq
10\log_{10}
\left(
\frac{\mathrm{SNR}_{\mathrm{ANF\text{-}DC}}^{\max}}
{\mathrm{SNR}_{\mathrm{ANF\text{-}AC}}^{\max}}
\right)
\\
&=
10\log_{10}
\left(
\frac{\pi y^\star}{2J^2(y^\star)}
\right) \approx \boxed{0.84~\mathrm{dB}}.
\end{aligned}
\label{eq:snr_gap_analog_digital}
\vspace{-1mm}
\end{equation}
Thus, under the adopted approximations, the presented \emph{analog} ANF--AC at its optimal Rx array size achieves an SNR within \emph{only $\approx$$0.84$\,dB of the maximum SNR with the digital ANF--DC}.

Importantly, the proposed ANF--DC closely matches the performance of the theoretically optimal MRC combiner~\cite{proakis2001digital}. Thus, the proposed ANF--AC, defined in \eqref{eq:anf_ac_combiner} and \eqref{eq:pfc_definition}, combined with optimal Rx size selection via \eqref{eq:optimal_antenna_product}, can bring the SNR with the analog Rx \emph{within less than $1$\,dB} of the best-possible fully digital MRC under the same distance, rotation, and the number of Tx and Rx antennas.

%%%%%%%%%%%%%%%%%%%%%%%%%%%%%%%%%%%%%%%%%%%%%%%%%%%%%%%%%%%%%%%%%%%%%%%%%%%%%%%%%%%%%%%%%%%%%%%%%%%%%%%%%%%%%%%%%%%%%%%%%%%%%%%%%%%%%%%
%%%%%%%%%%%%%%%%%%%%%%%%%%%%%%%%%%%%        Section V. Numerical Results        %%%%%%%%%%%%%%%%%%%%%%%%%%%%%%%%%%%%%%%%%%%%%%%%%%%%%%%
%%%%%%%%%%%%%%%%%%%%%%%%%%%%%%%%%%%%%%%%%%%%%%%%%%%%%%%%%%%%%%%%%%%%%%%%%%%%%%%%%%%%%%%%%%%%%%%%%%%%%%%%%%%%%%%%%%%%%%%%%%%%%%%%%%%%%%%
\section{Numerical Results}
\label{sec:results}
\vspace{+0.5mm}
We numerically elaborate our analytical results for a LoS link between Tx and Rx ULAs, possibly misaligned by $\alpha$ and $\theta$ (Fig.~\ref{fig:system_model}), at $f = 300$\,GHz. The Tx power spectral density is $P_\text{T}=0.1$\,W$/$$2.16$\,GHz, where $0.1$\,W = $20$\,dBm and $2.16$\,GHz is a unit channel width for millimeter-wave and sub-terahertz radios~\cite{ieee_802_15_3d_standard_2024}. Unless stated otherwise, we set $N_{\text{T}} = 1025$, $N_{\text{R}} = 127$, $N_0 = -174$\,dB/Hz, and the Tx--Rx separation $R = 10$\,m.

To validate the accuracy of our mathematical analysis (leveraging the approximate NF channel model and other approximations), \emph{all our numerical results} model the \emph{exact} channel from \eqref{eq:channel_coefficient} and \eqref{eq:element_distance} and not its approximation in Sec.~\ref{subsec:closed_form_effective_channel}.

%%%%%%%%%%%%%%%%%%%%
\vspace{-0.5mm}
\subsection{Performance of Proposed Combiners ANF--AC \& ANF--DC}
\label{subsec:proposed_combiner_performance}
\vspace{-0.5mm}
We start with Fig.~\ref{fig:downlink_combiner_performance}, which compares Rx combining strategies across Tx--Rx distances $R$ and ULA misalignment angles $\alpha$ and $\theta$. Across all the regimes in Fig.~\ref{fig:snr_vs_distance}--Fig.~\ref{fig:snr_vs_rotation}\footnote{In Fig.~\ref{fig:snr_vs_distance}, results are accurate for $R > R_{1} = 0.25 \sqrt[3]{N_{\text{T}}^{4}\lambda^{3}/2} \approx 2.05$\,m (Fresnel boundary, validity range of the array-type NF channel in Sec.~\ref{sec:system_model}~\cite{Selvan2017Fraunhofer}).}, the SNR with the ANF--AC closely matches the one with the theoretically optimal, non-analytical analog combiner EGC. ANF--DC SNRs match those with the MRC (SNR gap is~$\leq$$0.006$\,dB).

Hence, the two proposed analytical combiners provide useful benchmarks for future NF studies. Having established their close match to EGC and MRC, we omit the latter from subsequent figures and focus on ANF--AC (EGC-like) and ANF--DC (MRC-like). Both outperform FF beam steering, with SNR gains of up to $7.5$\,dB at $R$$=$$10$\,m. Also, Tx NF beamfocusing with ANF--DC approaches the SVD-MRC (Tx-side singular value decomposition, SVD~\cite{emil_book}), demonstrating the effectiveness of NF beamfocusing with digital combining.

Compared with analog focusing-back toward the Tx center, ANF--AC achieves higher SNR. As shown in Fig.~\ref{fig:snr_vs_distance}, the gain reaches $\approx$$5$\,dB at shorter distances but diminishes as $R$ approaches the Fraunhofer near-field boundary $R_{2} = \lambda (N_{\text{T}} + N_{\text{R}})^2 / 2$~\cite{cui2023nearfield}, where the FF plane-wave model reduces all considered combiners to beam steering.

In Fig.~\ref{fig:snr_vs_azimuth}, ANF--AC outperforms Rx beamfocusing by $\approx$$3$--$4$\,dB at zero azimuth, with the gain decreasing as $\lvert\alpha\rvert$ increases. Similarly, increasing $\lvert\theta\rvert$ in Fig.~\ref{fig:snr_vs_rotation} improves the SNR for all the schemes (as focused beams span more in distance than angle) but reduces the ANF--AC advantage.

%%%%%%%%%%%%%%%%%%%%
\subsection{Optimal Rx Size for Beamfocusing and Analog Combining}
\label{subsec:optimal_receiver_aperture}
\vspace{-0mm}
We now illustrate further applications of the analytical Rx combiner properties. The result in \eqref{eq:optimal_boundary_db} has already been illustrated in Fig.~\ref{fig:amplitude_profile} and discussed in Section~\ref{subsec:optimal_beam_capture_level}. We proceed with Fig.~\ref{fig:snr_vs_nr_comparison}, which examines the optimal Tx--Rx size trade-off in \eqref{eq:optimal_antenna_product} for downlink ($N_{\text{T}} = 1025$) and uplink ($N_{\text{T}} = 127$, corresponding to an $\approx$$6$\,cm-long array, fitting a tablet device).

In Fig.~\ref{fig:snr_vs_nr_comparison}, the SNR initially increases with $N_{\text{R}}$ for all schemes but then diverges. For digital combiners, it saturates at $N_{\text{R}}$$\gtrsim$$200$ in downlink and $N_{\text{R}}$$\gtrsim$$500$ in uplink, reaching $\approx$$45$\,dB, close to the maximum ANF--DC SNR in \eqref{eq:snr_max_analog_digital2}. Analog combiners, including ANF--AC, reach a peak and then degrade as the physical Rx aperture exceeds the Tx beam's focal spot.

Fig.~\ref{fig:snr_vs_nr_comparison} yields several others observations. \emph{First,} the optimal $N_{\text{R}}$ for ANF--AC using the exact NF channel closely matches the theoretical result in \eqref{eq:optimal_antenna_product}, derived for the approximate channel. \emph{Second,} at this optimum, ANF--AC outperforms both FF beam steering and the con-focusing approach in~\cite{mizmizi2026focusingconfocusing}, although con-focusing is more stable across $N_{\text{R}}$. \emph{Third,} for larger $N_{\text{R}}$, ANF--AC consistently outperforms Rx beamfocusing by exploiting the additional PFC coefficients in \eqref{eq:pfc_definition}. As Fig.~\ref{fig:snr_vs_nr} shows, ANF--AC also shifts the cross-over with FF beam steering toward larger $N_{\text{R}}$ and higher SNR. \emph{Finally,} across all modeled scenarios, the gap between the maximum SNRs with the analog ANF--AC and the digital ANF--DC remains $\approx$$0.84$,dB, confirming our SNR ratio derived in \eqref{eq:snr_gap_analog_digital}.

\begin{figure}[t]
    \centering
    \includegraphics[width=\linewidth]{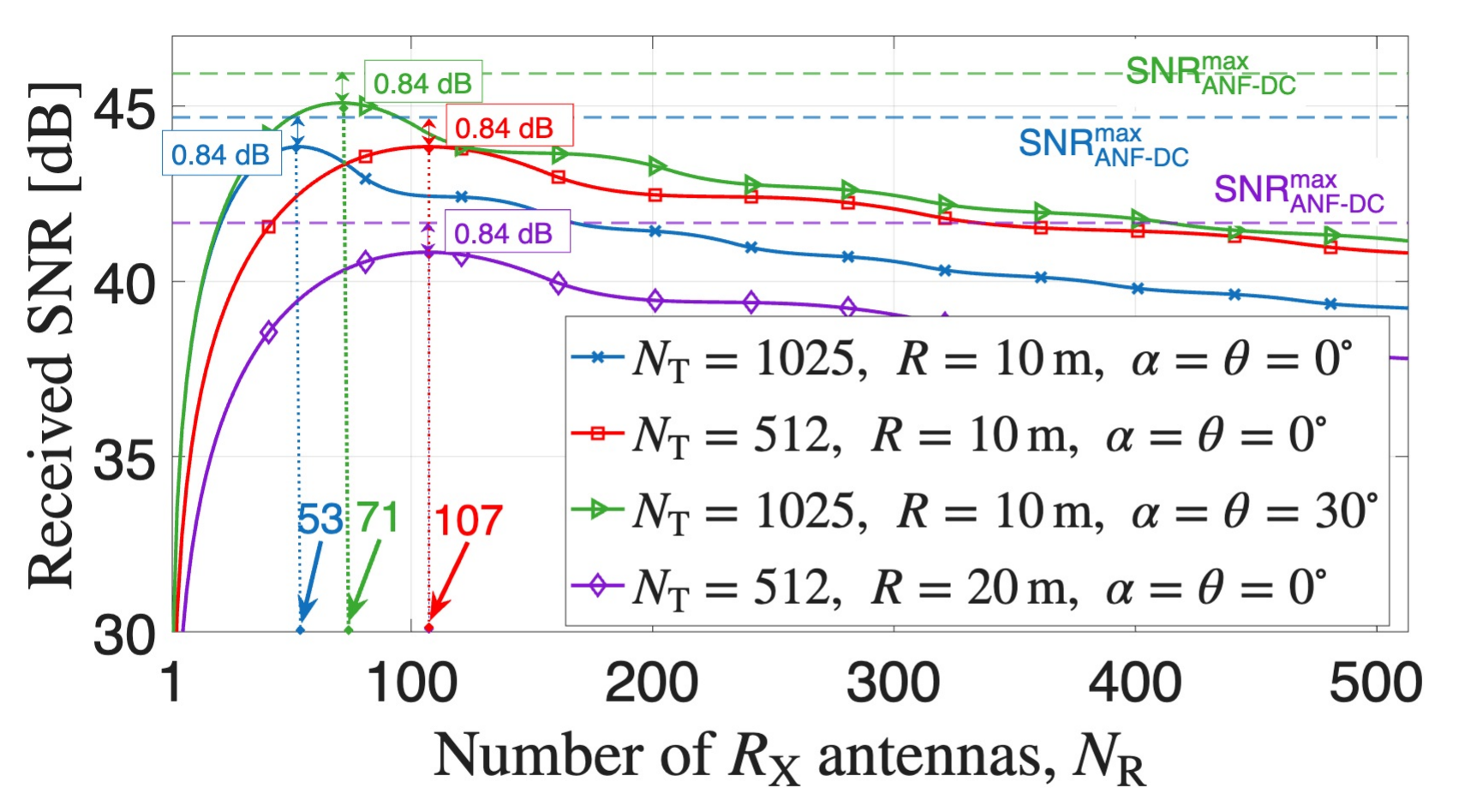}
    \vspace{-3mm}
    \caption{Received SNR variation with the ANF--AC combiner.}
    \label{fig:SNR_vs_Nr_variations}
    \vspace{-2mm}
\end{figure}

Fig.~\ref{fig:SNR_vs_Nr_variations} compares ANF--AC across Tx/Rx array sizes, Tx--Rx distances, and ULA misalignments. \emph{First,} shorter $R$ (e.g., $10$\,m vs. $20$\,m) favors smaller $N_{\text{R}}$ and yields higher SNR, owing to lower path loss and a tighter Tx focal spot. \emph{Second,} consistent with \eqref{eq:optimal_antenna_product}, larger $N_{\text{T}}$ favors smaller $N_{\text{R}}$. ULA misalignment can further improve SNR by a few dB, as also observed in Fig.~\ref{fig:snr_vs_azimuth} and Fig.~\ref{fig:snr_vs_rotation}. Importantly, the theoretical $0.84$\,dB SNR gap calculated in \eqref{eq:snr_gap_analog_digital} holds across all modeled configurations.

%%%%%%%%%%%%%%%%%%%%%%%%%%%%%%%%%%%%%%%%%%%%%%%%%%%%%%%%%%%%%%%%%%%%%%%%%%%%%%%%%%%%%%%%%%%%%%%%%%%%%%%%%%%%%%%%%%%%%%%%%%%%%%%%%%%%%%%
%%%%%%%%%%%%%%%%%%%%%%%%%%%%%%%%%%%%        Section VI. Conclusions        %%%%%%%%%%%%%%%%%%%%%%%%%%%%%%%%%%%%%%%%%%%%%%%%%%%%%%%%%%%%
%%%%%%%%%%%%%%%%%%%%%%%%%%%%%%%%%%%%%%%%%%%%%%%%%%%%%%%%%%%%%%%%%%%%%%%%%%%%%%%%%%%%%%%%%%%%%%%%%%%%%%%%%%%%%%%%%%%%%%%%%%%%%%%%%%%%%%%
%\balance
\section{Conclusions}
\label{sec:conclusion}
Most prior beamfocusing studies either assume an idealized point-source Rx, limiting practical relevance, or employ \emph{non-analytical} Rx combiners, such as EGC and MRC, hence challenging closed-form tractability. To address this gap, we proposed two \emph{analytical} combiners: ANF--AC and ANF--DC.

Both achieve near-optimal performance in the considered line-of-sight NF scenarios, with and without antenna array misalignment, incurring less than $0.006$\,dB loss relative to EGC and MRC, respectively. Then, by leveraging the analytical properties of the presented combiners, we: (i)~established in closed form the optimal Tx--Rx ULA size trade-off for maximizing the received SNR; and (ii)~derived that the SNR gap between an optimized analog Rx using ANF--AC and a fully digital Rx using optimal MRC does not exceed $0.84$\,dB.

Future work includes extending our analysis to uniform planar arrays and non-focusing near-field wavefronts, as well as utilizing the results for mathematical link- and system-level studies of coverage, capacity, reliability, interference, and energy efficiency in future near-field wireless systems.

%%%%%%%%%%%%%%%%%%%%%%%%%%%%%%%%%%%%%%%%%%%%%%%%%%%%%%%%%%%%%%%%%%%%%%%%%%%%%%%%%%%%%%%%%%%%%%%%%%%%%%%%%%%%%%%%%%%%%%%%%%%%%%%%%%%%%%%
%%%%%%%%%%%%%%%%%%%%%%%%%%%%%%%%%%%%%%%%%%    References    %%%%%%%%%%%%%%%%%%%%%%%%%%%%%%%%%%%%%%%%%%%%%%%%%%%%%%%%%%%%%%%%%%%%%%%%%%%
%%%%%%%%%%%%%%%%%%%%%%%%%%%%%%%%%%%%%%%%%%%%%%%%%%%%%%%%%%%%%%%%%%%%%%%%%%%%%%%%%%%%%%%%%%%%%%%%%%%%%%%%%%%%%%%%%%%%%%%%%%%%%%%%%%%%%%%
\bibliographystyle{IEEEtran}
\bibliography{references}

\end{document}